\documentclass[prb,twocolumn,,amsmath,amssymb,floatfix]{revtex4}
\usepackage{graphicx}
\usepackage{bm}

\def \be{\begin{equation}}
\def \ee{\end{equation}}
\def \ba{\begin{array}}
\def \ea{\end{array}}
\def \beq{\begin{eqnarray}}
\def \eeq{\end{eqnarray}}

\begin{document}
\title{Breakdown of the adiabatic limit in low dimensional gapless systems}

\author{Anatoli Polkovnikov$^1$ and Vladimir Gritsev$^2$}
\affiliation {$^1$Department of Physics, Boston University, Boston, MA 02215\\
$^2$Department of Physics, Harvard University, Cambridge, MA 02138 }

\begin{abstract}
It is generally believed that a generic system can be reversibly
transformed from one state into another by sufficiently slow change 
of parameters. A standard argument favoring this assertion is based 
on a possibility to expand the energy or the entropy of the system 
into the Taylor series in the ramp speed. Here we show that this 
argumentation is only valid in high enough dimensions and can break down 
in low-dimensional gapless systems. We identify three generic regimes of 
a system response to a slow ramp: (A) mean-field, (B) non-analytic, and 
(C) non-adiabatic. In the last regime the limits of the ramp speed going 
to zero and the system size going to infinity do not commute and the 
adiabatic process does not exist in the thermodynamic limit. We support 
our results by numerical simulations. Our findings can be relevant to 
condensed-matter, atomic physics, quantum computing, quantum optics, 
cosmology and others.

\end{abstract}

\maketitle

Adiabatic or reversible (also known as quasi-stationary) processes
by all means play a major role both in physics and technology. The
adiabatic process is formally defined as such where no heat is
transferred to the system from the environment~\cite{LL5}. Typically
such processes occur at time scales which are fast enough compared
to the thermalization times with the environment but yet which are
sufficiently slow so that the system always remains in thermal
equilibrium. Another feature of an adiabatic process is the
conservation of entropy. There is a simple general argument showing
that slow quasi-stationary processes are adiabatic and thus
reversible. The argument goes as follows:~\cite{LL5} Assume that in
an isolated system some external parameter $\kappa$ is slowly driven
from some initial value $\kappa_{\cal A}$ to the final one
$\kappa_{\cal B}$. For simplicity we assume that $\kappa$ changes
linearly in time, though it is not necessary, and let $\delta$ be
the rate of this change. We also refer to $\delta$ as to the ramp
speed. We assume that on the way the system does not undergo any
discontinuous phase transitions (though second order phase
transitions are generally allowed). Then the entropy density (or the
entropy per unit volume) of the system in the final state, $S_{\cal
B}$, will be a function of this parameter $\delta$ and we can expect
that for small enough $\delta$ one can expand $S_{\cal B}$ into the
Taylor series:
\be
S_{\cal B}=S_{\cal
A}+\alpha^\prime\delta+\beta^\prime\delta^2+\dots\, .
\ee
On general grounds we can argue that $\alpha^\prime\equiv 0$ because
the entropy can only increase and thus can not be sensitive to the
sign of $\delta$. Thus at small $\delta$ the excess entropy density
pumped into the system during this process is
\be
\Delta S_{\cal AB}\approx \beta^\prime \delta^2.
\label{S_B}
\ee
In other words it vanishes quadratically as $\delta\to 0$. We note
there are some subtleties with this expression at zero temperature,
which we will mention later.

The adiabatic theorem in thermodynamics is intimately related to the
adiabatic theorem in quantum mechanics, which states that under slow
enough external perturbations there are no transitions between
energy levels. Thus, for example, if one starts from a unique ground
state and adiabatically tunes the system into another regime of
parameters the system remains in the ground state and thus no
entropy is generated. The quantum mechanical adiabatic theorem
implies thermodynamic adiabatic theorem, however, there are some
subtleties involved~\cite{footnote}. If there is a gap in the system
then at zero temperature the ground state can be excited only
through the so called Landau Zener mechanism~\cite{LZ}, which gives
the exponentially small probability of excitations and thus the
exponentially small entropy increase. We note that the fact that
this increase is exponential rather than quadratic is a peculiarity
of the zero temperature limit. However, systems which have a gap are
rather exception than the rule. Indeed most of systems with broken
continuous symmetries have gapless excitations over the ground state
(Goldstone modes). Thus solids have phonons (sound waves),
ferromagnets and anti-ferromagnets have gapless magnon or spin-wave
excitations, superfluids have gapless Bogoliubov excitations and so
on. Even in gapped systems, for example superconductors, one always
has continuum spectrum above the gap, which is occupied by
quasi-particles at finite temperatures. If gapless excitations are
present in the system then the Landau Zener mechanism does not help
to protect against creating excitations. However, one can generally
argue that the available phase space for the excitations decreases
as the adiabaticity parameter $\delta$ becomes smaller and one can
still expect that Eq.~(\ref{S_B}) holds.

The adiabatic theorem can be also formulated for integrable systems
(the simplest example of these are noninteracting systems described
by harmonic theories). Then the entropy is no longer a good
observable because integrable systems do not thermalize. Instead one
can use the excitation (quasi-particle) density $n_{\rm ex}$. The
quantum adiabatic theorem implies that $n_{\rm ex}$ does not change
if the process is sufficiently slow. Then on general grounds we can
expect that the analogue of Eq.~(\ref{S_B}) will be $\Delta n_{\rm
ex}(\delta)\approx\tilde\beta \delta^2$. Whether the system is
integrable or not, its energy is always a good observable, since it
is defined for any state of the system in or out of equilibrium. In
terms of energies the adiabatic theorem can be formulated in the
same spirit:
\be
\mathcal E_{\mathcal B}(\delta)=\mathcal E_{\mathcal
B}(0)+\beta\delta^2,
\label{E_B}
\ee
where $\mathcal E_{\mathcal B}(0)$ is the energy of the state
adiabatically connected to the initial state.

There is one important caveat in the considerations described above.
They are similar in spirit to the mean-field argument suggesting
existence of the {\it long range order} in systems with broken
continuous symmetries. However, we know that at low dimensions
either quantum~\cite{subir} or thermal~\cite{chaikin lubenskiy}
fluctuations can change the picture and destroy the long range order
completely. An ultimate reason for this is that at low dimensions
the density of the low energy states $\rho(\epsilon)$ is generically
high. Indeed if $\epsilon(q)\propto q^z$, where $q$ is the momentum
of an excitation and $\epsilon(q)$ is its energy, then we
necessarily have $\rho(\epsilon)\propto \epsilon^{d/z-1}$, where $d$
is the dimensionality of the system. The quantum or thermal
fluctuations force excitations to occupy these low energy states and
if their density is sufficiently high, they can qualitatively change
the nature of the equilibrium state. The same arguments can be
applied to the adiabatic process we are interested in. Indeed no
matter how slow the ramp is (in a gapless system) there will be
always created some low energy excitations. In low dimensions these
excitations can significantly alter behavior of the system and in
particular invalidate Eqs.~(\ref{S_B}, {\ref{E_B}}). In the
remainder of the paper we will address this issue in detail.

The main conclusion of our work is that there are three possible
regimes of a system response to a slow process: ({\bf A})
mean-field, ({\bf B}) non-analytic, and ({\bf C}) non-adiabatic. In
the first regime ({\bf A}) we find that indeed the energy of the
system and other thermodynamic quantities are analytic functions of
the adiabaticity parameter $\delta$ and Eq.~(\ref{E_B}) is valid.
This regime is always realized at sufficiently high dimensions. In
the second regime ({\bf B}) one still has the adiabatic limit in the
sense that there is a well defined limit $\delta\to 0$ but the
correction in $\delta$ to the energy density or other quantities
behaves nonanalytically:
\be
\mathcal E_{\cal B}(\delta)\approx \mathcal E_{\cal B}(0)+\beta
|\delta|^\nu,
\ee
where $\nu<2$ is some power which depends on the dimensionality and
on the universal details of the spectrum. And finally the third
({\bf C}) non-adiabatic regime is most unusual. There instead of
Eq.~(\ref{E_B}) we have
\be
\mathcal E_{\cal B}(\delta)\approx \mathcal E_{\cal B}(0)+\beta
|\delta|^\nu L^\eta,
\ee
where $L$ is the system size and $\eta>0$ is another exponent. In
this regime there is no adiabatic process in the thermodynamic
limit. In other words the limits $\delta\to 0$ and $L\to\infty$ {\it
do not commute}. This is a striking conclusion, which states that no
matter how slowly we try to drive the system, if the latter is
sufficiently large, we will never reach the adiabatic limit. As we
show both analytically and numerically such regime is realized in
(but not limited to) the situation where one starts from an ensemble
of noninteracting bosons in one and two dimensions at finite
temperature and slowly increases the interaction strength. We point
that existence of the regime {\bf C} is consistent with recent
theoretical~\cite{pokrovsky, gurarie} and experimental~\cite{regal}
works suggesting that there is no adiabatic limit in a particular
problem of a BCS-BEC crossover in cold atoms.

Let us make another very important remark. A general argument
predicting entropy conservation in a slow process assumes that the
system is always in thermal equilibrium. If, on the other hand, we
are dealing with an integrable or a nearly integrable system then
thermalization times can be either infinite or very long. However,
as we argued above the non-integrability of the system and its
thermalization are not necessary conditions for formulation of the
adiabatic theorem in the generalized sense. The only genuine
requirement is that the system is initially in a stable equilibrium.
In this work we rigorously study the limit where the system either
does not thermalize or it rethermalizes only after the ramp and show
that all three regimes are possible. We support our results by
performing numerical simulations for a particular interacting
non-integrable system and show that the effects of weak
nonintegrability {\it do not affect} our conclusions.

An alternative point of view on the existence of the regimes {\bf B}
and {\bf C} is the breakdown of the linear response theory for slow
perturbations~\cite{ehud}. In this work we concentrated on the
spatially uniform case. However, one can anticipate that similar
breakdown of the linear response will be relevant to a more general
class of nonuniform low-frequency perturbations. We are leaving the
detailed analysis of such possibilities for a future work.

Apart from many applications in condensed matter and atomic physics
our findings may be relevant to such diverse fields as quantum
optics, quantum computing, in particular for adiabatic quantum
computation~\cite{FGGS}; inflationary cosmology, which has the phase
of adiabatic evolution of Gaussian scale-invariant quantum
fluctuations in the slow-roll approximation~\cite{Li}; cosmic
microwave background radiation~\cite{Parker}, and Hawking radiation
of black holes~\cite{Hawking}. In all these fields the assumption of
adiabaticity in slow processes is usually taken for granted, while
our results suggest that this might not be always the case.

{\em Analytical treatment.} In this paper we consider a specific low
energy quadratic Hamiltonian:
\be
\mathcal H=\sum_q {\rho_s q^2\over 2}|\phi_q|^2+{1\over
2}\kappa_q\Pi_q^2,
\label{Hamiltonian}
\ee
where $\phi_q$ and $\Pi_q$ are the coordinate and the conjugate
momentum. We note that this Hamiltonian describes a very wide class
of gapless systems. Thus in solids $\phi_q$ represents the phonon or
the plasmon field, in ferromagnets and anti-ferromagnets $\phi_q$
describes magnons or spin-waves, in superfluids $\phi_q$ describes
Bogoliubov's excitations. This list can be easily extended further.
Depending on the system, the couplings $\rho_s$ and $\kappa_q$ have
different meanings. For example, for superfluids $\rho_s$ denotes
the superfluid density, and $\kappa_q$ is related to the
compressibility. Dependence of $\kappa_q$ on $q$ also varies for
different systems. Thus $\kappa_q={\rm const}(q)$ in solids,
anti-ferromagnets and superfluids and $\kappa_q\propto q^2$ is
ferromagnets and noninteracting Bose systems. The reason why the
Hamiltonian~(\ref{Hamiltonian}) is so generic and applicable to such
different situations is that it describes Goldstone modes of systems
with a broken continuous symmetry. We further choose
$\kappa_q=\kappa+\lambda q^2$. This choice allows us to cover all
the situations mentioned above. In a superfluid $\kappa$ stands for
the compressibility of the system and we will use this terminology
in the paper.

Let us now imagine that one slowly increases $\kappa$ in time
\be
\kappa(t)=\kappa_0+\delta t,
\ee
where $\kappa_0$ is the initial value of the compressibility. In
principle, one can consider ramps of other parameters, but we do not
expect any significant changes in the overall picture. In this paper
we will focus on positive sign of $\delta$, i. e. the situation
where compressibility increases in time. It can be shown that the
opposite process, where $\kappa$ decreases, gives similar results up
to unimportant numerical prefactors. We also comment that the choice
$\kappa_q(t)=\kappa(t)+\lambda q^2$ allows us to analyze interesting
possibilities from the field theory point of view. Indeed formally
$\lambda q^2$ is an irrelevant coupling, which is unimportant at low
energies (this simply follows from the fact that it scales to zero
at small $q$). On the other hand, if one starts at zero or very
small $\kappa_0$ this term dominates the behavior of the system at
initial times and as we will see it qualitatively changes the
system's response. Thus $\lambda$ plays the role of a dangerously
irrelevant variable~\cite{subir} in our problem. On the other hand
if $\kappa_0$ is large then indeed one can safely set $\lambda$ to
zero.

Let us first assume that the system is initially prepared in the
ground state, i.e. consider the zero temperature limit. Since the
Hamiltonian (\ref{Hamiltonian}) is noninteracting it does not lead
to thermalization, therefore as we discussed the entropy is not a
good concept. One can instead describe the degree of
non-adiabaticity of the system by computing either the density of
excitations $n_{\rm ex}$ created during the ramp or the energy
density (energy per unit volume) pumped to the system $\mathcal E$.
Since our model is noninteracting, the evolution of the wave
function (or more generally density matrix) can be explicitly
obtained. We give details of such analysis in the section
``Methods''. Here we will outline the main results.

Let us first assume that the system is initially prepared at zero
temperature. Then its wave function factorizes to the product of
Gaussians corresponding to the ground state of a harmonic oscillator
for each momentum $q$ (see Eq.~(\ref{psi})). We find there are two
different regimes of the system behavior depending on whether the
initial compressibility $\kappa_0$ is finite or zero. The crossover
between the two regimes occurs at $\kappa_0^\star\approx
\delta\sqrt{\lambda/\rho_s}$. Note that as $\delta\to 0$ we have
$\kappa_0^\star\to 0$. For large initial compressibility,
$\kappa_0\gg\kappa_0^\star$, the response of the system belongs to
the {\bf A} regime in all three spatial dimensions, i.e. the energy
density behaves as $\mathcal E\propto \delta^2$. Actually in one
dimension there is an additional logarithmic correction to this
scaling and $\Delta\mathcal E\propto \delta^2|\ln\delta|$. If
$\kappa_0=0$ then the {\bf B} regime is realized in all three
spatial dimensions and the energy density scales as
\be
\mathcal E\propto {\delta^{(d+1)/4}\over
(\rho_s\lambda^3)^{(d+1)/8}}.
\label{en_0}
\ee

We note that if one analyzes the density of excitations $n_{\rm ex}$
then the classification of different regimes is different. In
particular for large $\kappa_0$ one finds that the nonanalytic {\bf
B} regime is realized in one spatial dimension and the mean-field
{\bf A} regime occurs above two dimensions. For $\kappa_0=0$ the
system is in the nonadiabatic {\bf C} regime for $d=1$ and it is in
{\bf B} regime in two and three dimensions. The fact that the
classification of the system's response according to the density of
excitations is different from that based on the energy is not really
surprising. Indeed primarily low energy modes are occupied and they
do not contribute much to the total energy. A similar effect also
occurs in equilibrium.

If the original system is nonintegrable then it will rethermalize at
long times. Thermalization in closed nonintegrable systems is a
separate issue, which we are going to address in future work. We
just point out that the Liouville's theorem stating that the entropy
of an isolated system is conserved in time is generally not an issue
in non-integrable many-particle systems. Indeed the amount of
external noise needed to break this entropy conservation scales to
zero exponentially with the number of particles so that
non-integrable systems are never completely isolated. Let us assume
that after the ramp we can wait long enough so that the system does
rethermalize and let us see what are the consequences for the
entropy. Note that since the system is isolated from the
environment, thermalization occurs at a fixed energy, which we just
determined above. This energy thus should be related to the
temperature of the final state $T_{f}$ via
\be
\mathcal E=C_d {T_{\rm f}^{d+1}\over (\kappa_{\rm f}\rho_s)^{d/2}},
\label{eq:42}
\ee
where $C_d$ is a numerical constant. Equating Eqs.~(\ref{en_0}) and
(\ref{eq:42}) we can find that in the case of $\kappa_0=0$
\be
T_{\rm f}\propto \sqrt{\kappa_{\rm f}}\delta^{1/4}{\rho_s^{3/8}\over
\lambda^{3/8}}, \Rightarrow \mathcal S_{\rm f}=
A_d{\delta^{d/4}\over (\rho_s\lambda^3)^{d/8}}.
\label{eq:44}
\ee
We see that the behavior of the entropy essentially follows that of
the energy with a slightly different exponent. We note that the fact
that exponent is different is a peculiarity of the zero temperature
case, where the entropy has a singular limit. Apart from that the
same classification of different regimes {\bf A}, {\bf B}, and {\bf
C} applies to the entropy of the system, its temperature and other
thermodynamic quantities.

Quite similarly one can analyze the response of the system to the
ramp if it is initially prepared at some finite temperature $T$.
Because at finite temperatures the excitations are present in the
system from the beginning, we will be interested in excess
quantities pumped to the system during the ramp like $\Delta n_{\rm
ex}$, $\Delta \mathcal E$, and $\Delta S$. The analysis of finite
temperature evolution is straightforward and we sketch it in Section
``Methods'' and Appendix~\ref{App:B}.

The results are again strongly sensitive to the initial
compressibility. Thus if $\kappa_0$ is large then in one dimension
we get
\be
\Delta\mathcal  E=A {|\delta|T\sqrt{\kappa_f}\over
\sqrt{\rho_s}\kappa_0^2}
\ee
In turn, if the system rethermalizes after the ramp, this results in
the nonanalytic correction to the entropy:
\be
\Delta S\approx {dS\over d\mathcal E}\Delta \mathcal E= A
{|\delta|\over \sqrt{\rho_s}\kappa_0^{3/2}}.
\ee
In this case obviously the nonanalytic {\bf B} regime is realized.
In two and three dimensions $\Delta \mathcal E\propto \delta^2$
(with logarithmic corrections at $d=2$) and thus we have the mean
field {\bf A} regime.

If $\kappa_0=0$ then in one and two dimensions the energy density
diverges with the system size:
\be
\Delta \mathcal E=A_d {T\sqrt{\kappa_f}\over \lambda\rho_s^{1/6}}
\delta^{1/3}L^{7/3-d}
\label{en_1d}
\ee
and thus
\be
\Delta S=A_d^\prime \left({T\over
\lambda\rho_s^{2/3}}\right)^{d\over d+1}\delta^{d\over
3(d+1)}L^{(7-3d)d\over 3 (d+1)}.
\ee
In particular at $d=1$ we have $\Delta S\propto \delta^{1/6}L^{2/3}$
and at $d=2$: $\Delta S\propto (\delta L)^{2/9}$. So in one and two
dimensions the behavior of the system is nonadiabatic and the {\bf
C} regime is realized. In three dimensions there is no dependence on
the system size and we find
\be
\Delta\mathcal E=A_3
{T\sqrt{\kappa_f}\over\rho_s^{1/4}\lambda^{5/4}}\sqrt{\delta}\quad
{\rm and}\quad \Delta S=A_3^\prime {\sqrt{T}\over
\sqrt{\rho_s}\lambda}\sqrt{\delta}.
\ee
So that the system is in the non-analytic {\bf B} regime. We note
again that the density of excitation diverges stronger than the
energy. In particular at finite $T$ and for $\kappa_0=0$ we find
$\Delta n_{\rm ex}\propto \delta^{1/3} L^{10/3-d}$, i.e. it diverges
with the system size in all three spatial dimensions.

{\em Numerical results: application to interacting bosons.} While
the analysis of the previous sections is formally exact, it directly
applies only to a special case of an integrable harmonic system.
Most of the real systems are nonintegrable. However, at low energies
the harmonic approximation is usually very good. We already argued
that whether the system is allowed to re-thermalize at long times or
not, the qualitative picture does not change. If the thermalization
occurs then a good measure of the heating is the entropy generated
during the ramp, if not then one should look into the generated
density of excitations. Thermalization in a closed system is
certainly a very interesting and not very well understood problem,
which however requires a separate analysis and goes beyond the scope
of this paper. So we will focus on the analysis of the energy
(density) pumped to the system, $\Delta\mathcal E$, since this
observable is not sensitive to the details of thermalization.

To perform numerical simulations we choose the Bose-Hubbard model on
a square lattice described by the Hamiltonian:
\be
\mathcal H_{bh}=-J\sum_{\langle ij\rangle} (a_i^\dagger
a_j+a_j^\dagger a_i)+ {U(t)\over 2}\sum_j a_j^\dagger
a_j(a_j^\dagger a_j-1),
\label{h_bh}
\ee
where $a_j$ and $a_j^\dagger$ are the annihilation and creation
operators of bosons on the j-th site, $J$ represents the tunneling
matrix element and $U$ is the interactions strength. The sum in the
first term is taken over the nearest neighbor pairs. We take $J$ to
be time independent and the interaction increasing in time according
to $U(t)=U_0\tanh(\delta t)$. So $U(t)$ first increases linearly in
time and then saturates at some steady state value. For small enough
interactions $U_0\ll J n_0$, where $n_0$ is the mean number of atoms
per lattice site, in the quadratic approximation the Hamiltonian
(\ref{h_bh}) maps to the Bogoliubov Hamiltonian (\ref{Hamiltonian})
with $\rho_s\approx 2J n_0$, $\kappa\approx U$, and $\lambda\approx
\rho_s/4n_0^2$. In order to simulate dynamics of the system we
employ the semiclassical approach developed by one of us~\cite{twa}.
In this approach one expands the time evolution of the system in the
small quantum parameter $U/Jn_0$.  We note for those more familiar
with the Keldysh technique~\cite{kamenev} that this approach treats
all classical vertexes exactly and expands the evolution in quantum
vertexes. In the leading order in this parameter one obtains the so
called truncated Wigner approximation (TWA)~\cite{walls, steel},
where the classical fields $\psi_j^\star$ and $\psi_j$ corresponding
to operators $a_j^\dagger$ and $a_j$ satisfy the time dependent
Gross-Pitaevskii equations of motion. In the next order the
classical fields are subject to a single quantum jump during the
evolution. We find that while TWA approximation is adequate at
finite temperatures, in the zero temperature limit one has to go
beyond and add the next correction. This finding agrees with a
general statement that the semiclassical approximation can break
down at long times~\cite{twa, psg}. We present some details of the
semiclassical approach in Appendix~\ref{App:C} and here only will
discuss the results.

First we look into one dimensional systems, since there, according
to the theory, we expect strongest effects of nonadiabaticity. In
order to avoid potential complications related to strong quantum
effects we choose the parameters of the system deep in the
superfluid regime throughout the entire evolution: $n_0=20, J=1,
U_0=0.25$ so that the semiclassical parameter $U_0/Jn_0\sim 10^{-2}$
(see Ref.~[\onlinecite{psg}]) is very small. Note that despite
$U_0=0.25$ is relatively small the product $U_0n_0=5$ is larger than
$J=1$ implying that at long times $U(t)\approx U_0$ the system is in
so called quantum rotor regime with healing length smaller than the
lattice spacing. According to our theoretical expectations the
system is mostly excited while the healing length remains large, i.
e. when $U\ll U_0$. In this regime $U(t)=U_0\tanh\delta t\approx
U_0\delta t$ linearly increases with $t$ thus we can directly
compare numerical results with the analytical predictions of the
previous section.

\begin{figure}[ht]
\includegraphics[width=8.5cm]{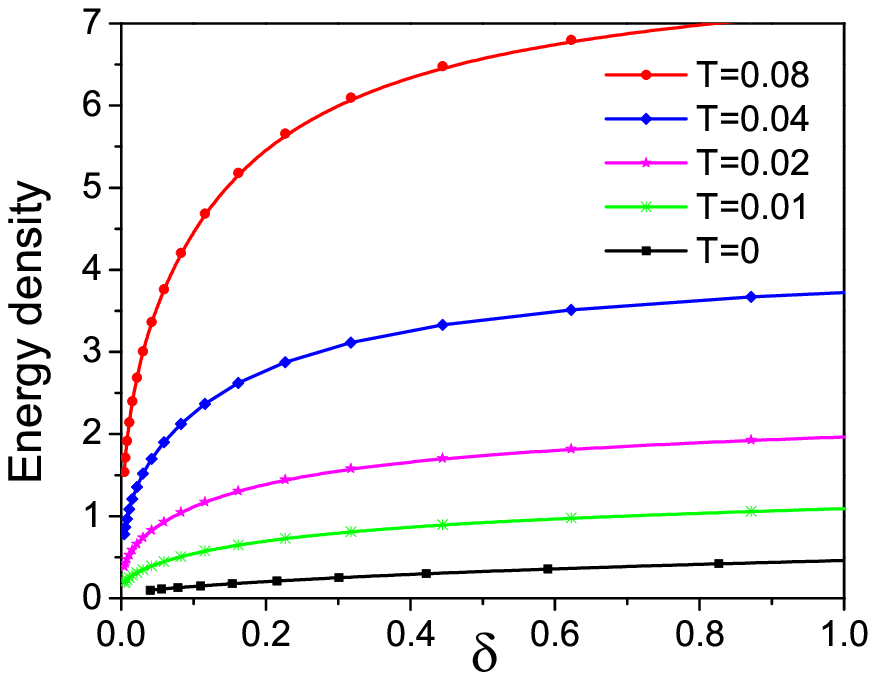}
\includegraphics[width=8.5cm]{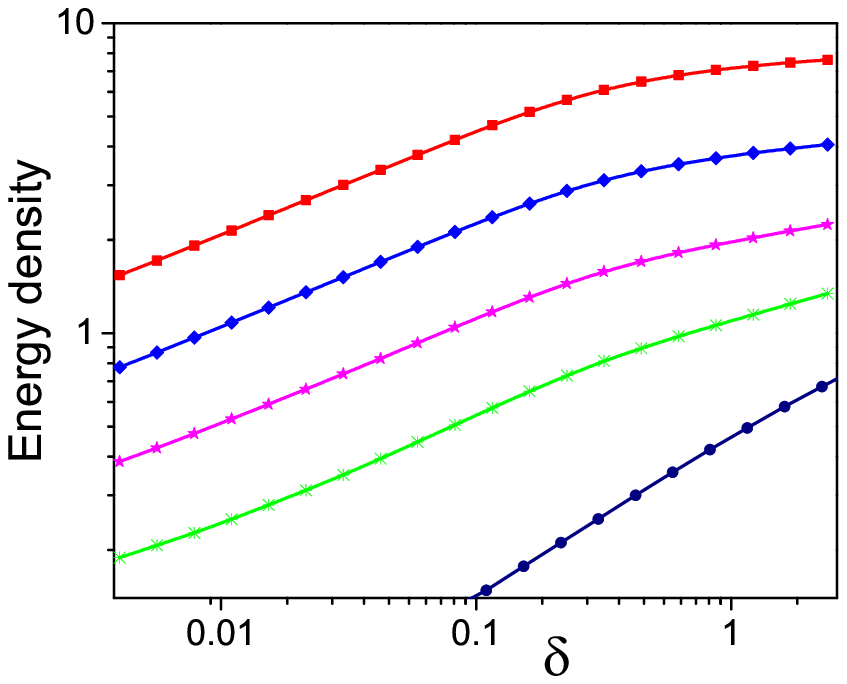}
\caption{ Dependence of the energy density pumped into the system
during the ramp $\Delta \mathcal E$ on the parameter $\delta$ for
different initial temperatures and a fixed system size $L=128$. The
other parameters are $J=1$, $n_0=20$, and $U_0=0.25$. The
interaction increases in time according to $U(t)=U_0\tanh \delta t$.
The top graph is in the normal scale and the bottom graph is in the
log-log scale. At finite temperatures the dependence of
$\Delta\mathcal E$ on $\delta$ agrees with the prediction of
Eq.~(\ref{en_1d}): $\Delta\mathcal E\propto \sqrt[3]\delta$ (the
power $1/3$ can be read from the slope of the dependence
$\Delta\mathcal E(\delta)$ in the lower graph). Similarly at zero
temperature the behavior agrees with Eq.~(\ref{en_0}):
$\Delta\mathcal E\propto \sqrt{\delta}$. Note also that at small
ramps the curves at different temperatures are equidistant
indicating that $\delta\mathcal E\propto T$ again in accord with
Eq.~(\ref{en_1d}).}
\label{fig:finte_temp}
\end{figure}

\begin{figure}[ht]
\includegraphics[width=8.5cm]{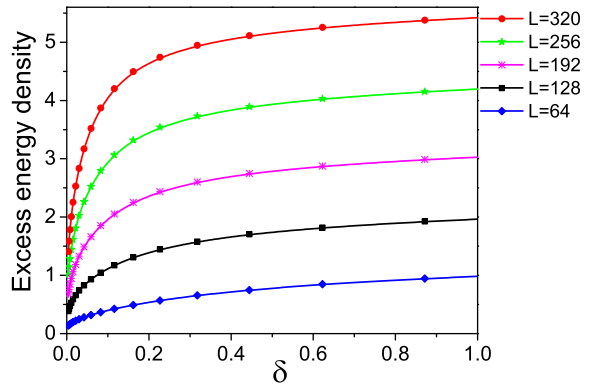}
\caption{Dependence of $\Delta \mathcal E$ on $\delta$ for different
sizes at fixed temperature $T=0.02$. As we argued in the previous
section (see Eq.~(\ref{en_1d})), there is clearly no thermodynamic
limit and the heating becomes more severe with the growth of the
system size.}
\label{fig:finte_temp2}
\end{figure}

In Fig.~\ref{fig:finte_temp} we show the dependence of the energy
per site pumped into the system during the ramp as a function of the
parameter $\delta$ for different temperatures $T$ at a fixed system
size $L=128$. Note that even at very low temperatures $T\sim 0.01J$
the behavior of $\Delta \mathcal E$ is dominated by the thermal
effects. This trend becomes even more apparent if we analyze
dependence of $\Delta \mathcal E$ on $L$ (see
Fig.~\ref{fig:finte_temp2}).
In agreement with the analytic results $\Delta \mathcal E$ strongly
grows with the system size (see Eq.~(\ref{en_1d})) and clearly there
is no thermodynamic adiabatic limit. One can also check that the
dependence of $\Delta \mathcal E$ on $L$ agrees with the prediction
of Eq.~(\ref{en_1d}). Although we do not show the graphs but we
checked that at zero temperature $\Delta \mathcal E$ is best fitted
by $\sqrt{\delta}$ dependence and is almost insensitive to the
system size, again in agreement with the analytic
results~(\ref{en_0}).
\begin{figure}[ht]
\includegraphics[width=8.5cm]{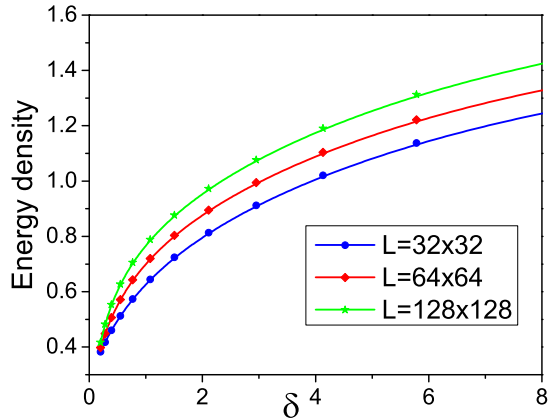}
\caption{Dependence of $\Delta \mathcal E$ on $\delta$ in
two-dimensional system for two different sizes. The temperature is
fixed at $T=0.2$, the other parameters are the same as in
Fig.~\ref{fig:finte_temp2}. The asymptotical behavior of $\Delta
\mathcal E$ at small $\delta$ agrees with $\delta^{1/3}$ dependence
predicted by Eq.~(\ref{en_1d}). Also $\Delta \mathcal E$ slowly
increases with the system size consistent with the analytic
prediction $\Delta\mathcal E\propto L^{1/3}$.}
\label{fig:finte_temp3}
\end{figure}

We also performed numerical simulations for the two-dimensional
Bose-Hubbard model. We work with similar parameters as in the
one-dimensional case, except that at much higher temperature $T=0.2$
and slightly smaller linear sizes. The reason we had to choose
higher $T$ is that at a given temperature thermal effects in two
dimensions dominate the system behavior at much larger linear sizes
than in one dimension. We find that the results are again in a good
agreement with predictions of Eq.~(\ref{en_1d}). In particular, we
find that $\Delta \mathcal E\propto \delta^{1/3}$ at small $\delta$
and that $\Delta \mathcal E$ slowly grows with the system size
consistent with $\Delta\mathcal E\propto L^{1/3}$.

{\em Summary and outlook.} In this paper we analyzed possible
breakdown of the standard adiabatic (or quasi-stationary)
approximation in low-dimensional gapless systems. We explicitly
analyzed behavior of harmonic systems which are described by
quadratic phonon-like excitations. Despite this particular choice,
we point again that the Hamiltonian~(\ref{Hamiltonian}) describes a
wide class of phenomena such as phonons in solids, magnons in ferro-
and antiferromagnets, Bogoliubov excitations in superfluids and so
force. Using both analytical and numerical methods we showed that
generically one can have three possible heating regimes. The first
{\bf A} regime is mean-field like. There one can apply simple
arguments showing that various thermodynamic observables are
analytic functions of the ramp speed $\delta$. This regime is
typically realized at high dimensions. In the second regime {\bf B},
which we called non-analytic, the energy, entropy and other
quantities depend on $\delta$ in a nonanalytic way: $\mathcal
E_{\mathcal B}(\delta)\approx \mathcal E_{\mathcal B}(0)+\beta
|\delta|^\nu$, where $\nu<2$. The exponent $\nu$ depends on the
universal critical exponents characterizing the gapless phase. And
finally in low dimensions one can have a very unusual regime {\bf
C}, which we called nonadibatic. In this regime the limits of
$\delta\to 0$ and the system size $L\to\infty$ do not commute and
for example the energy density behaves as $\mathcal E_{\mathcal
B}(\delta)\approx \mathcal E_{\mathcal B}(0)+\beta
|\delta|^{\nu}L^\eta$. So no matter how slowly one ramps the system,
in the thermodynamic limit one can never reach the adiabatic regime.

We did not attempt to classify all possible models and model
situations where different regimes can be realized. This is probably
a very difficult task. But as a matter of principle we proved that
all three regimes are possible. We point that there is an
interesting connection between this work and some earlier works
where slow dynamics across a quantum critical point was studied. See
Refs.~[{\onlinecite{adiabatic, zurek1, jacek, fubini}}]. In
particular, using perturbative approach, which we also discuss in
Appendix~\ref{APP:Pert}, one of us showed~\cite{adiabatic} that the
density of excitations generated during this process scales as
\be
\Delta n_{\rm ex}\propto \delta^{d\nu/(z\nu+1)},
\label{qpt}
\ee
where $z$ and $\nu$ are the dynamical and correlation length
exponent characterizing the phase transition. These results were
later confirmed by exact methods~\cite{jacek, fubini}. But this is
nothing but the {\bf B} regime of heating, which we proposed in this
paper. And indeed in Ref.~[\onlinecite{adiabatic}] it was shown that
the scaling (\ref{qpt}) is valid below some critical dimension,
where the exponent of $\delta$ saturates at $2$ and we are back to
the {\bf A} regime. We emphasize that our general arguments
presented here do not exclude a possibility of crossing a continuous
second order phase transition during the ramp. For example in
Refs.~[\onlinecite{zurek1, jacek, fubini, jacek1, caneva}], which
considered crossings of a critical point in various integrable spin
chains, the scaling~(\ref{qpt}) was attributed to the Kibble-Zurek
mechanism~\cite{kibble zurek}. However, we stress that neither the
derivation of this scaling nor the analyzed systems do not involve
any topological defects. These works rather showed that the
considered problems belong to the nonanalytic regime {\bf B} of
non-adiabaticity. The fact that Eq.~(\ref{qpt}) is correct is the
consequence of the validity of the perturbation theory. As we showed
in this paper the perturbative expression can break down due to the
bunching effect if the excitations near the critical point have
bosonic nature. Then the scaling (\ref{qpt}) will no longer be
valid. From results of this paper on general grounds one can also
expect stronger dependence of $\Delta n_{\rm ex}$ on $\delta$ and
possible emergence of the non-adiabatic {\bf C} regime at finite
temperatures.

An obvious outcome of our analysis is that one has to be very
careful with making statements about adiabaticity in isolated or
nearly isolated gapless low-dimensional systems. This can be
important in many various situations ranging from realizing
proposals on adiabatic quantum computation and preparation of
interacting systems in a given state via slow ramps to inflationary
cosmology and black hole radiation. Perhaps cold atoms are the
systems where our results can be immediately tested in experiments.
There one has all the necessary experimental tools like isolation
from the environment together with high tunability of parameters of
the system and the possibility to perform tuning in real
time~\cite{bloch_review}. However, as coherent dynamic becomes more
and more important in other fields like nano-physics, quantum
optics, cavity QED, etc., we expect that our results should be
relevant to more and more potential applications.

\section*{METHODS.}

{\em Zero temperature.} The Hamiltonian (\ref{Hamiltonian}) is
quadratic and thus the time evolution can be found exactly. We first
look into the zero temperature case. For quadratic Hamiltonians it
is well known that if the initial wave function is gaussian, it will
remain gaussian for an arbitrary time dependence of the parameters
of the Hamiltonian. Since we assume that the system was originally
prepared in the ground state the initial wave function is
\be
\Psi(\{\phi_q\})=\prod_q {1\over
(2\pi\sigma_{0,\,q})^{1/4}}\exp\left[-{\phi_q^2\over
4\,\sigma_{0,\,q}}\right],
\label{psi}
\ee
where $ \sigma_{0,\,q}=1/ (2q)\sqrt{\kappa_{0,q}/\rho_s }$. If
$\kappa$ changes with time, $\sigma_q$ acquires time dependence:
\be
i {d \sigma_q\over dt}=2\rho_s q^2\sigma_q^2-{1\over 2}\kappa_q(t).
\label{eq:16}
\ee
This is a Riccati equation, which can be explicitly solved through
Airy functions. We give details of this solution in the
Appendix~\ref{App:A}.

It is straightforward to check that the number of excitations per
mode $q$ is related to $\sigma_q$ via:
\be
n_q={1\over 2}\left[{\sigma^{\rm eff}_q\over \sigma^{\rm
eq}_q}-1\right],
\label{n_q}
\ee
where $\sigma^{\rm eff}_q=1/\Re(\sigma_q^{-1})$ and $\sigma^{\rm
eq}_q$ is the equilibrium (ground state) value of $\sigma$ in the
final state. We point that at large values of $\kappa$ the ratio
$\sigma^{\rm eff}_q/\sigma^{\rm eq}_q$ does not depend on $\kappa$
(see details in the Appendix~\ref{App:A}). The asymptotical
expressions for $n_q$ in the limit of large and small $q$ can be
easily found from Eq.~(\ref{n_q}) and the solution of
Eq.~(\ref{eq:16}) presented in Appendix~\ref{App:A}. In particular,
\be
n_q\approx {\delta^2\over 64 q^2\rho_s \kappa_{0,\,q}^{\,3}}
\label{n_q1}
\ee
at $q\sqrt{\rho_s\kappa_{0,\,q}^{\,3}}\gg\delta$ and
\be
n_q\approx {\pi\over 3^{2/3}\Gamma^2(1/3)}{\delta^{1/3}\over
q^{1/3}(\rho_s\kappa_{0,\,q}^{\,3})^{1/6}}
\label{n_q2}
\ee
in the opposite limit, where $\Gamma(x)$ stands for the
gamma-function. We note that the asymptotics (\ref{n_q1}) exactly
coincides with the result of the perturbation theory, which can be
obtained from Eq.~(\ref{eq:3}).

The total density of excitations $n_{\rm ex}$ can be obtained by
integrating $n_q$ over $q$. Let us consider two different cases. If
$\kappa_0\gg \delta\sqrt{\lambda/\rho_s}$ then $n_{\rm ex}$ is
dominated by the integral of Eq.~(\ref{n_q1}) with the cutoff
$q\sim\delta$ and we immediately recover the perturbative result, i.
e. $n_{\rm ex}\propto |\delta|^d$ for $d<2$ and $n_{\rm ex}\propto
\delta^2$ for $d>2$ (in two dimensions $n_{\rm ex}\propto \delta^2
|\ln\delta|$). So we recover that again only type {\bf A} and type
{\bf B} regimes are possible. The situation, where one starts in the
noninteracting regime is more complicated. Indeed Eq.~(\ref{n_q2})
indicates that the number of excited modes at small $q$ diverges in
this case as $q^{-4/3}$. In two dimensions and above this is an
integrable divergence and we again recover the perturbative result
$n_{\rm ex}\propto \delta^{d/4}$. However in one dimension this
integral diverges at small $q$ and thus should be formally cutoff at
$q\sim 1/L$. In this case one finds that $n_{\rm ex}\propto
\delta^{1/3}L^{1/3}$. This is precisely the regime {\bf C}, which
was missing in the perturbative analysis. We already highlighted
that the reason why the perturbation theory fails is the Bose
enhancement of the transition probabilities due to their bunching
tendency.

Once we know the complete wave function we can, in principle, find
arbitrary observables like various correlation functions, response
to external probes etc. However, one has to realize that finding
wave function was possible only due to exact integrability of the
harmonic theory. So one should look into quantities, which are
robust to effects of weak nonintegrability. One of such quantities,
which has a particular importance in thermodynamics is the energy
density of the system:
\be
\mathcal E\approx {\sqrt{\rho_s\kappa_{\rm f}}\over 2}\int {d^d
q\over (2\pi)^d}\, q\left({\sigma^{\rm eff}_q\over \sigma^{\rm
eq}_q}-1\right),
\label{eq:41}
\ee
where $\kappa_{\rm f}$ is the final compressibility. The integral
above converges in all dimensions leading to Eq.~(\ref{en_0}).

{\em Finite temperature case.} One can generalize the previous
results to the situation where the system is initially prepared in
the thermal state characterized by some temperature $T$. In this
case the excitations are present in the system from the beginning
and we will be interested in analyzing their enhancement during the
ramp. Instead of the wave function we have to deal with the density
matrix, but essentially the derivation is similar to the zero
temperature case. We present the details of the analysis in the
Appendix~\ref{App:B}. The result of these calculations is
surprisingly simple:
\be
\left. {\sigma^{\rm eff}_q\over \sigma^{\rm
eq}_q}\right|_T=r_q\left.{\sigma^{\rm eff}_q\over \sigma^{\rm
eq}_q}\right|_{T=0},
\ee
where
\be
r_q=\coth\left[{q\sqrt{\kappa_{0,\,q} \rho_s}\over 2 T}\right].
\ee
In other words one needs to take zero temperature asymptotics for
the width of the wave function and multiply them by $r_q$. In the
zero temperature limit $r_q\equiv 1$ and we obviously reproduce the
results at $T=0$. At high temperatures $T\gg
q\sqrt{\rho_s\kappa_{0,\, q}}$ we have
\be
r_q\approx {2T\over q\sqrt{\rho_s \kappa_{0,\,q}}}
\ee
and thus $\sigma^{\rm eff}_q$ diverges at $q\to 0$ much faster than
in the zero temperature case. Let us observe that the number of
excitations of a $q$-mode is
\be
n_q={1\over 2}\left[{\sigma^{\rm eff}_q\over \sigma^{\rm
eq}_q}-1\right]=r_q \left.n_q\right|_{T=0}+{1\over 2}(r_q-1).
\ee
The last term is nothing but the equilibrium occupation of the
$q$-mode at the initial time. Therefore we see that multiplying the
zero temperature result for $n_q$ by $r_q$ we immediately obtain the
number of additional excitations in the $q$-mode $\Delta n_q$
generated during the ramp.

Using Eqs.~(\ref{n_q1}) and (\ref{n_q2}) at
$q\sqrt{\rho_s\kappa_{0,\,q}^{\, 3}}\gg\delta$ we get
\be
\Delta n_q\approx {\delta^2 T\over
32\,q^3\rho_s^{3/2}\kappa_{0,\,q}^{7/2}}
\ee
and in the opposite limit
\be
\Delta n_q\approx {2\pi\over
3^{2/3}\Gamma^2(1/3)}{\delta^{1/3}T\over
q^{4/3}\rho_s^{2/3}\kappa_{0,\,q}}.
\ee
Now even at finite value of $\kappa_0$ the total density of
generated excitations $\Delta n_{\rm ex}$ diverges in one dimension
$\Delta n_{\rm ex}\propto \delta^{1/3}L^{1/3}$ and we are in the
non-adiabatic {\bf C} regime. We note that formally $n_{\rm ex}$
also diverges at finite temperature at $d=1$ even in equilibrium,
but this divergence is much weaker and scales as the logarithm of
the system size. In two dimensions we have $\Delta n_{\rm ex}\propto
|\delta|$ and thus the nonanalytic {\bf B} regime is realized. And
finally in three dimensions and above $\Delta n_{\rm ex}\propto
\delta^2$ and thus the mean field {\bf A} regime works (in three
dimensions there are logarithmic corrections to $\delta^2$ scaling).
If the initial state is noninteracting then the divergence of $n_q$
at small $q$ is much more severe: $n_q\propto q^{-10/3}$ and in all
three spatial dimensions the non-adiabatic {\bf C} regime is
realized:
\be
\Delta n_{\rm ex}=A_d {\delta^{1/3}T\over
\rho_s^{2/3}\lambda}L^{1/3+3-d}.
\ee
Thus in order to keep the excitation density constant one has to
scale $\delta$ as $L^{3d-10}$. This means, for example, that in one
dimension if the system size increases by a factor of two, then in
order to keep heating ($\Delta n_{ex}$) the same one has to decrease
$\delta$ by a factor of $128$!

{\em Acknowledgements} We would like to acknowledge  E.~Altman, E.
Demler, S.~Girvin, V. Gurarie, M.~Lukin, V.~Pokrovsky, and N.
Prokof'ev for useful discussions. A.P. was supported by AFOSR YIP and partially
by NSF under Grant PHY05-51164. V. G. is partially supported by the
Swiss National Science Foundation and AFOSR. A.P. also acknowledges
Kavli Institute for Theoretical Physics for hospitality.

\appendix

\section*{Supplementary Material}

\section{Fermi golden rule analysis}
\label{APP:Pert}

The easiest way to find the density of excitations $n_{\rm ex}$
produced during a slow increase of $\kappa(t)$ in
Hamiltonian~(\ref{Hamiltonian}) is to use Fermi golden rule
analysis. Indeed, since we expect that $n_{\rm ex}$ is small at
small $\delta$ one can expect that the perturbation theory in
$\delta$ should give a good estimate of $n_{\rm ex}(\delta)$ at
small $\delta$. Using a general expression for the density of
excitations derived in Ref.~[\onlinecite{adiabatic}] we find:

%
\be
n_{ex}={1\over 32}\int {d^d q\over
(2\pi)^d}\left|\int_{\kappa_{0,q}}^\infty {d\xi\over\xi}\exp\left(
{4i\over 3\delta}\sqrt{\rho_s}q\,\xi^{3/2}\right)\right|^2,
\label{eq:3}
\ee
where $\kappa_{0,\,q}=\kappa_0+\lambda q^2$. This expression gives
different asymptotics in two opposite limits. (i) If $\delta\gg
\kappa_0^2\sqrt{\rho_s/\lambda}$, which is the case if one starts in
from weak interactions $\kappa_0\to 0$, then
\be
n_{ex}\approx A_d {\delta^{d/4}\over
\rho_s^{d/8}\lambda_{\phantom{s}}^{3d/8}},
\label{eq:7}
\ee
where $A_d$ is a numerical constant. It is easy to check that above
$8$ dimensions the exponent of $\delta$ saturates at $2$ and does
not further change with the dimensionality. Expression (\ref{eq:7})
suggests that in this particular situation the nonanalytic regime
{\bf B} is realized in all physical dimensions. In one dimension it
is particularly hard to reach the adiabatic regime since $n_{\rm
ex}$ scales only as $\delta^{1/4}$.

We note that the scaling in Eq.~(\ref{eq:7}) is consistent with the
one obtained in Ref.~[\onlinecite{adiabatic}] for the crossing of
the second order phase transition: $n_{\rm ex}\propto
\delta^{d\nu/(z\nu+1)}$, where $\nu$ is the critical exponent
characterizing divergence of the correlation length. In our case
there is a diverging healing length $\xi\sim \sqrt{\lambda/\kappa}$
instead of the correlation length (see Ref.~[\onlinecite{ps}] for
details) so that $\nu=1/2$ and given that $z=2$ in the
noninteracting regime one immediately recovers that
$\nu/(z\nu+1)=1/4$.

In the opposite limit (ii) where the initial value of $\kappa$ is
large $\delta\ll\kappa_0 \sqrt{\rho_s/\lambda}$ the situation
becomes more diverse. Thus for dimensions $d<2$ Eq.~(\ref{eq:7})
yields
\be
n_{ex}\approx A_d^\prime {\delta^d\over
\rho_s^{d/2}\kappa_0^{3d/2}}.
\label{n_ex}
\ee
On the other hand for $d>2$ the exponent saturates and we have
\be
n_{ex}\approx A_d^\prime {\lambda^{1-d/2}\kappa_0^d\over
\rho_s}\,\delta^2.
\label{eq:11}
\ee
In two dimensions there is an additional logarithmic correction to
the scaling (\ref{n_ex}). We see that in this situation the critical
dimension above which the mean field regime holds is $d^\star=2$.

The present analysis can be generalized to other situations. For
example, in the case of ferromagnets $\kappa_0\equiv 0$ and then one
can tune $\lambda$. Then one finds that $n_{\rm ex}\propto
\delta^{d/2}$ and the critical dimension is $d^\star=4$. We comment
that one can also consider other scenarios of varying $\kappa$ with
time. For example, if $\kappa\propto (\delta t)^r$ then it is easy
to see that $n_{\rm ex}\propto \delta^{dr/2(r+1)}$. As $r$ increases
the scaling of the density of excitations interpolates from
$\delta^{d/4}$ to $\delta^{d/2}$ and changes $d^\star$ from $8$ to
$4$.

This simple perturbative analysis shows the existence of {\bf A}
({\bf B}) regimes for dimensions above (below) some critical value
$d^\star$. This analysis, however, misses the existence of the {\bf
C} regime. To justify the validity of the application of the Fermi
golden rule one has to require that the probability of excitation of
each momentum mode is small. This requirement breaks down at low
energies as can be readily seen from Eq.~(\ref{eq:3}). In the case
when the excitations have Fermionic character, which is e.g. the
case for crossing the critical point in the transverse field Ising
model or the XXZ chain~\cite{subir}, the mistake of the perturbative
treatment is a simple factor of the order of one (see
Refs.~[\onlinecite{adiabatic, zurek1, jacek, fubini}]). The
Goldstone modes described by the Hamiltonian~(\ref{Hamiltonian}) on
the other hand are harmonic oscillators and thus behave as bosons.
Bosons unlike fermions have a bunching tendency, i. e. transition
probabilities can be significantly enhanced compared to the golden
rule prediction.

\section{Evolution of the wave function under the ramp at zero
temperature}
\label{App:A}

Here we present some details of the solution of Eq.~(\ref{eq:16}).
For convenience we write it here again:
\be
i {d \sigma_q\over dt}=2\rho_s q^2\sigma_q^2-{1\over 2}\kappa_q(t).
\label{eq:16a}
\ee

This equation can be simplified by first changing independent
variable $t$ to $\kappa_q(t)$ and then by simple rescaling:
\be
\kappa=\tilde\kappa {\delta^{2/3}\over \sqrt[3]{\rho_s q^2}},\;
\sigma_q=\tilde\sigma_q {\delta^{1/3}\over 2\sqrt[3]{\rho_s^2
q^4}},\; q=\tilde q {\delta^{1/4}\over \rho_s^{1/8}\lambda^{3/8}}.
\label{eq:18a}
\ee
Under these rescalings we also have $\tilde
\kappa_q=\tilde\kappa+\tilde q^{8/3}$. Then one can check that
Eq.~(\ref{eq:16}) is equivalent to
\be
i{d\tilde \sigma_q\over d\tilde \kappa_q}=\tilde \sigma_q^2-\tilde
\kappa_q.
\label{eq:18}
\ee
This Riccati equation can be explicitly solved in terms of Airy
functions ${\rm Ai}$ and ${\rm Bi}$:
\be
\tilde \sigma_q=-i {{\rm Bi}^\prime (-\tilde \kappa_q)+\alpha_q {\rm
Ai}^\prime (-\tilde \kappa_q)\over {\rm Bi}(-\tilde
\kappa_q)+\alpha_q {\rm Ai}(-\tilde \kappa_q)},
\ee
where $\alpha_q$ is an integration constant, which is determined
from the initial conditions. In the limit $\tilde \kappa_q\to
\infty$ ignoring unimportant fast oscillating terms we find
\be
\Re \left[1\over \tilde\sigma_q\right]\to {2\Im\alpha_q\over
\sqrt{\tilde \kappa_q}[1+|\alpha_q|^2]}.
\ee
Note that the real part of $1/\sigma_q$ determines $|\psi|^2$ and
thus the probability distribution of the corresponding phase. The
fact that $1/\sigma_q\to 0$ as $\tilde \kappa_q\to\infty$ should not
be surprising. Indeed the width of the ground state wave function in
scaled variables is
\be
\sigma^{\rm eq}_q=\sqrt{\tilde \kappa_q}\approx {1\over
2q}\sqrt{\kappa\over\rho_s}.
\ee
The probability of excitations in the system is determined by the
ratio of $\sigma_q$ and $\sigma^{\rm eq}$, which takes a well
defined limit at $\kappa\to\infty$. Introducing $\sigma^{\rm
eff}_q=1/\Re(\sigma_q^{-1})$ we find
\be
{\sigma^{\rm eff}_q\over\sigma^{\rm eq}_q}={1+|\alpha_q|^2\over
2\Im\alpha_q}.
\label{sig_eff}
\ee
The initial condition determining $\alpha$ is:
\be
\sqrt{\tilde \kappa_{0,\,q}}=i { {\rm Bi}^\prime(-\tilde
\kappa_{0,\,q})+\alpha_q {\rm Ai}^\prime(-\tilde
\kappa_{0,\,q})\over {\rm Bi}(-\tilde \kappa_{0\,q})+\alpha_q {\rm
Ai}(-\tilde \kappa_{0,\,q})}.
\ee
This equation can be inverted to give
\be
\alpha_q=-{\sqrt{\tilde \kappa_{0,\,q}}\, {\rm Bi}(-\tilde
\kappa_{0,\,q})-i {\rm Bi}^\prime(-\tilde \kappa_{0,\,q})\over
\sqrt{\tilde \kappa_{0,\,q}}\, {\rm Ai}(-\tilde \kappa^q_0)-i {\rm
Ai}^\prime(-\tilde \kappa_{0,\,q})}.
\label{alpha_q}
\ee

In the limit $\tilde\kappa_{0,\,q}\ll 1$ these equation yields:
\be
\alpha_q\approx \sqrt{3}+i{3^{2/3}\Gamma^2(1/3)\over
\pi}\sqrt{\tilde \kappa_{0,\,q}}.
\ee
Consequently
\be
{\sigma^{\rm eff}_q\over\sigma^{\rm eq}_q}\approx {2\pi\over
3^{2/3}\Gamma^2(1/3)}{1\over \sqrt{\tilde \kappa_{0,\,q}}}.
\label{eq:26}
\ee
In the opposite limit $\tilde\kappa_0^q\gg 1$ one finds
$\alpha_q\approx i$ and
\be
{\sigma^{\rm eff}_q\over\sigma^{\rm eq}_q}\approx 1+{1\over 32
\tilde \kappa_{0,\,q}^{\,3}}.
\label{eq:28}
\ee

\section{Evolution of the density matrix at finite temperatures.}
\label{App:B}

We choose to represent the density matrix corresponding to the
initial thermal state in the Wigner form~\cite{gardiner, walls}. For
the harmonic system described by the Hamiltonian~(\ref{Hamiltonian})
one can show that this density matrix factorizes into the product of
Gaussians:
\be
W_0=\prod_q {1\over 4\pi r_q}\exp\left[-{|\phi_{0,\, q}|^2\over
4\sigma_{0,\,q}r_q}-{\sigma_{0,\,q}|\Pi_{0,q}|^2\over 4 r_q}\right],
\label{eq:45}
\ee
where
\be
r_q=\coth\left[{q\sqrt{\kappa_{0,\,q} \rho_s}\over 2 T}]\right].
\ee
It is well known that in the noninteracting problem the time
evolution of the fields $\phi_{\bf q}$ and $\Pi_{\bf q}$ is
described by the classical equations of motion~\cite{gardiner}. In
particular,
\be
{d\over dt}\left[{1\over \kappa_q}{d\phi_q\over dt}\right]+\rho_s
q^2\phi=0,
\label{eq:47}
\ee
subject to initial conditions
\be
\phi_q(t=0)=\phi_{0,\, q},\; \dot\phi_q(t=0)=\kappa_{0\,q} \Pi_{0,\,
q},
\ee
where $\phi_{0, q}$ and $\Pi_{0, q}$ are randomly distributed
according to (\ref{eq:45}). The other important feature of Gaussian
ensembles is that in the absence of interactions the Wigner
distribution (\ref{eq:45}) always preserves its Gaussian form.
Therefore finding $\langle \phi_q^2(t) \rangle$ and $\langle
\Pi_q^2(t)\rangle$ is sufficient to fix the whole distribution
function at arbitrary times. Alternatively one can directly solve
the Liouville equation for the density matrix in the Wigner
form~\cite{gardiner} and come to the same conclusion.

A general solution of Eq.~(\ref{eq:47}) is:
\be
\phi_q(\tilde \kappa_q)=C_1 {\rm Ai}^\prime(-\tilde
\kappa_q)+C_2{\rm Bi}^\prime (-\tilde \kappa_q),
\ee
where as in Appendix~\ref{App:A} we changed the variables from $t$
to $\tilde \kappa_q$. The integration constants $C_1$ and $C_2$ can
be found from the initial conditions:
\beq
&&C_1={\pi \kappa_{0,\,q}\over \tilde
\kappa_{0,\,q}^{\,2}}{d\phi_{0,\,q}\over
d\kappa_{0,\,q}}{\rm Bi}^\prime(-\tilde \kappa_{0,\, q})
-\pi\phi_{0,\,q} {\rm Bi}(-\tilde \kappa_{0,\,q}),\phantom{XX}\\
&&C_2=\pi\phi_{0,\,q} {\rm Ai}(-\tilde \kappa_{0,\,q})-{\pi
\kappa_{0,\,q}\over \tilde \kappa_{0,\, q}^{\,2}}{d\phi_{0,\,q}\over
d\kappa_{0,\, q}}{\rm Ai}^\prime(-\tilde \kappa_{0,\,q}).
\eeq
From these expressions it is easy to find the asymptotical behavior
of $\langle \phi_q^2\rangle$ at large $\tilde \kappa$ and thus find
the width of the distribution $\sigma^{\rm eff}_q$:
\beq
&&{\sigma^{\rm eff}_q\over \sigma^{\rm eq}_q}={\pi\over 2} {r_q\over
\sqrt{\tilde \kappa_{0,\, q}}}\bigl[\tilde \kappa_{0,\,q}{\rm
Bi}^2(-\tilde \kappa_{0,\,q})+\tilde
\kappa_{0,\,q}{\rm Ai}^2(-\tilde \kappa_{0,\, q})\nonumber\\
&&~~~~~~~~~~~~~~~~~+({\rm Bi}^\prime(-\tilde \kappa_{0,\,
q}))^2+({\rm Ai}^\prime(-\tilde \kappa_{0,\,q}))^2\bigr].
\eeq
It is straightforward to verify that in the zero temperature limit
$r_q\to 1$ this expression coincides with the previous result (see
Eqs.~(\ref{sig_eff}) and (\ref{alpha_q})).

\section{Quantum expansion of the dynamics of a Bose-Hubbard
system.}
\label{App:C}

\begin{figure}[ht]
\includegraphics[width=8cm]{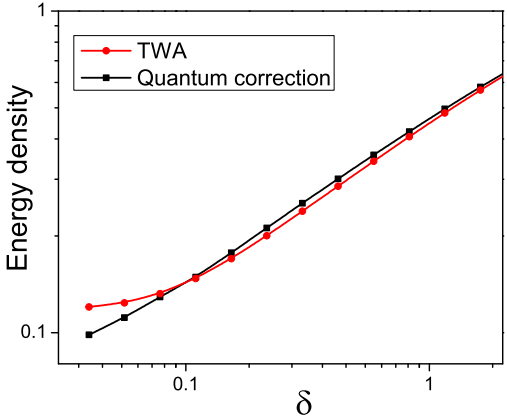}
\caption{ Dependence of the energy density  $\Delta \mathcal E$ on
the $\delta$ at zero temperature with and without the quantum
correction (\ref{quant_corr}). For the details of the calculation
and the parameters of the problem see Fig.~\ref{fig:finte_temp}.
Obviously at small values of $\delta$ the TWA breaks down and one
has to include the correction (\ref{quant_corr}).}
\label{fig:twa_quant}
\end{figure}

In the classical limit the bosonic fields $\psi_j^\star$ and
$\psi_j$ satisfy discrete Gross-Pitaevskii equations:
\be
i{\partial \psi_j\over \partial t}=-J\sum_{i\in O_j} \psi_i+U(t)
|\psi_j^2|\psi_j.
\ee
Here the sum in the first term is taken over the nearest neighbors
of the site $j$. In the leading order in quantum fluctuations,so
called truncated Wigner approximation (TWA), the fields $\psi_j$ and
$\psi_j^\star$ are subject to random initial conditions, which are
distributed according to the Wigner transform of the initial density
matrix $W(\psi_j^\star,\psi_j)$. The expectation value of an
arbitrary observable $\Omega(a_j^\dagger, a_j)$ is given by the
average of the corresponding Weyl symbol (fully symmetrized form of
the operator) $\Omega_{\rm cl}(\psi_j^\star,\psi_j)$ on the
solutions of the Gross-Pitaevskii equations:

\be
\langle \Omega(t)\rangle_0=\int D\psi^\star_jD\psi_j
W(\psi_j^\star,\psi_j)\Omega_{\rm cl}(\psi_j^\star(t)\psi_j(t)).
\ee

We find that this level of approximation gives very accurate results
in most of our simulations described in this paper. However, at zero
temperature case it breaks down for very slow ramps and we had to
include the next quantum correction to the TWA. The latter manifests
itself in the form of a single infinitesimal quantum jump during the
evolution:
\be
\psi_i(t^\prime)\to \psi_i(t^\prime)+\epsilon_1+i\epsilon_2.
\ee
The quantum correction is the evaluated as a nonlinear response of
$\Omega_{\rm cl}$ to such a jump~\cite{twa}:
\begin{widetext}
\be
\langle \Omega(t)\rangle_1=-\int D\psi^\star_jD\psi_j
W(\psi_j^\star,\psi_j)\sum_i \int_0^t dt^\prime {U(t^\prime)\over
16}\left[\Im\psi_i(t^\prime)
{\partial\over\partial\epsilon_1}-\Re\psi_i(t^\prime)
{\partial\over\partial\epsilon_2 }\right]\left[{\partial^2\over
\partial\epsilon_1^2}+{\partial^2\over
\partial\epsilon_2^2}\right]\Omega_{\rm
cl}(\psi_j^\star(t),\psi_j(t),\epsilon_1,\epsilon_2).
\label{quant_corr}
\ee
\end{widetext}

Numerically both the leading term $\langle \Omega(t)\rangle_0$ and
the next correction $\langle \Omega(t)\rangle_1$ are evaluated using
Monte-Carlo integration schemes. The third order derivatives in
Eq.~(\ref{quant_corr}) are found using finite differences, e. g.
\beq
&&{\partial^3 \Omega(\epsilon_1)\over\partial \epsilon_1^3}\approx
{\Omega(2\epsilon_1)-\Omega(-2\epsilon_1)-2\Omega(\epsilon_1)+2\Omega(\epsilon_1)\over
2\epsilon_1^3}\\
&& {\partial^3 \Omega(\epsilon_1,\epsilon_2)\over\partial
\epsilon_1\partial\epsilon_2^2}
 \approx {1\over 2\epsilon_1\epsilon_2^2 }\biggl(\Omega(\epsilon_1,\epsilon_2)+\Omega(\epsilon_1,-\epsilon_2)\\
 &&-\Omega(-\epsilon_1,\epsilon_2)
-\Omega(-\epsilon_1,-\epsilon_2)-2\Omega(\epsilon_1,0)+2\Omega(-\epsilon_1,0)\biggr).\nonumber
\eeq

It is easy to convince oneself that in order to evaluate these
finite differences one has to simultaneously solve $13$
Gross-Pitaevskii equations, one for $\epsilon_1=\epsilon_2=0$ and
the others for various combinations of $\epsilon_1, \epsilon_2=0,
\pm\epsilon, \pm 2\epsilon$. While solving 13 Gross-Pitaevskii
equations is certainly more time consuming task than solving one, it
is still tremendously more advantageous than dealing with the exact
quantum problem. To illustrate the importance of quantum correction
at zero temperature we show comparison of dependence $\Delta
\mathcal E(\delta)$ at zero temperature with and without this
correction (see Fig.~\ref{fig:twa_quant}).

Since the initial system is noninteracting, it is straightforward to
find the Wigner transform of the density matrix at finite
temperature $T$. It is more convenient to write it in the Fourier
space
\be
W(\hat\psi_k^\star,\hat\psi_k)=Z\prod_q
\exp\left[-2|\hat\psi_q|^2\tanh\left(\epsilon_0(q)-\mu\over
2T\right)\right],
\ee
where $\hat\psi_k$ is the discrete Fourier transform of $\psi_j$,
$Z$ is the normalization constant, $\epsilon_0(q)=-J\sum_{j}\mathrm
e^{iqj}$ is the excitation energy of the Bose-Hubbard
Hamiltonian~(\ref{h_bh}) in the absence of interactions and the
summation is taken over nearest neighbors of site at the origin,
$\mu$ is the chemical potential which enforces mean number of
particles per site $n_0$. We note that in large systems we consider
here, there is no difference in time evolution between grand
canonical and canonical ensembles~\cite{ap_cat}.


\begin{thebibliography}{99}
\bibitem{footnote} We would like to stress that in
this paper we understand the adiabatic theorem in the thermodynamic
sense, i.e. that the intensive quantities like density of excited
states scales to zero with $\delta$. There is a substantial
literature analyzing adiabaticity in the absolute sense, i.e. that
the systems remains in exact many body ground state (see
e.g.~[\onlinecite{FGGS}]). Such strong adiabaticity is always broken
in the thermodynamic limit even in gapped systems.
\end{thebibliography}
\end{document}